# Setting the stage: Building and maintaining a habitable world and the early conditions that could favor life's beginnings on Earth and beyond

Christopher K Jones[1], Michaela Leung[1], Chenyi Tu[1], Saleheh Ebadirad[1], Nate Marshall[1], Lin Tan[2], and Tim Lyons[1]

1. Department of Earth & Planetary Sciences, University of California, Riverside, CA 92521, USA
2. Department of Environmental Sciences, University of California, Riverside, CA 92521, USA

**Abstract**

The Hadean, once thought to be uninhabitable and tumultuous, has more recently been recontextualized as a clement time in which oceans, land, and life likely appeared on Earth. This non-exhaustive chapter follows multiple threads from planet formation to the origin of life. We place significant emphasis on the solar system context for the Earth, the timing and nature of crustal formation and the evolution of the surface and atmosphere. Several scenarios for prebiotic chemistry are also discussed including atmospheric photochemistry, wet-dry and freeze-thaw cycles, and hydrothermal vent systems. We attempt to draw connections between the large-scale, planetary processes and various origin of life pathways to illustrate possible overlaps and correlations. In detail, we conclude with and discuss the "impact of impacts" to show how asteroid and comet impacts during the Hadean may have affected many of these processes and scenarios, from generating land to altering the chemical composition and oxidation state of the early Earth's atmosphere and surface.

**Introduction**

The Archean, the topic of this volume, represents the first part of Earth's history with a significant rock record. It is also likely when life proliferated to the point of influencing biogeochemical cycles. The most notable example of this phenomenon is the rise of atmospheric oxygen observed at the end of the Archean and into the Proterozoic, commonly termed the Great Oxidation Event (GOE, 2.5–2.4 to 2.1–2.0 Ga) (Holland, 2002; Lyons et al., 2024). We refer to the roughly 500-600 million years before the Archean as the Hadean, because the Earth was historically imagined as being hot and uninhabitable at this time, thus drawing on the imagery of the Greek pantheon's god of the underworld: Hades. However, some lines of evidence imply that the early Earth environment was more temperate and conducive to the origin and sustainment of life, perhaps only a few hundred million years or less after the moon-forming impact (Carlson et al. 2014). A volume devoted to life and environments during the Archean is certain to benefit from a discussion of what came immediately before, and that is the mission of this contribution.

The timing, location, and mechanisms of life's origin are vigorously debated, with a wide diversity of proposed solutions ranging from reducing atmospheric chemistry to freeze-thaw cycles to hydrothermal vents (Miller and Urey, 1959; Martin et al., 2008; Damer and Deamer, 2020; Marín-Yaseli et al., 2020; Marchi et al., 2021). In each case, understanding the pathways for prebiotic

chemistry—that is, the abiotic synthesis of organic molecules that could become proto-cells and life over time—is critical, but remains elusive. For life as we know it, this transition from abiotic to biotic seems to fundamentally require a source of reduced carbon alongside gradients in temperature, pH, and redox, regardless of the host environment, to drive anabolic reactions toward the assembly of complex organic molecules (Baross and Hoffman, 1985; Macleod et al., 1994; Russell and Hall, 1997; Morse and Mackenzie, 1998; Stüeken et al., 2013; Harrison and Lane, 2018; Kitadai and Maruyama, 2018).

Much of the uncertainty surrounding the origin of life stems from debate about the composition of the early Earth's atmosphere and whether it was reducing or oxidizing. Arguments such as Miller and Urey (1959) invoke reducing atmospheric chemistry to synthesize organics, which can rain out of the atmosphere, accumulate on subaerial land surfaces, and polymerize through wet-dry or freeze-thaw cycles (Benner et al., 2020; Frenkel-Pinter et al., 2019; Frenkel-Pinter et al., 2022). However, studies based on analyses of the rare earth elements as a proxy for the mantle quartz-fayalite-magnetite buffer indicate that Earth's earliest atmosphere was likely not reducing and had minor amounts of $CH_4$ relative to $CO_2$, $N_2$, and $H_2O$ (Trail et al., 2011; Trail et al., 2012). These uncertainties warrant further investigations into the Hadean and the origins of life on Earth.

This text reviews several proposed prebiotic chemical pathways for the origins of life in the Hadean. We place these pathways in the context of large-scale planetary and environmental parameters such as the existence of tectonic plates, Earth's position in the solar system, the role of moons, and the chemistry of the oceans and atmosphere. To do this, we touch on critical themes that cross disciplines, including microbiology, atmospheric chemistry, geochemistry, and planetary science. We provide a non-exhaustive assessment of progress in understanding the relationships between life's beginnings and the processes, places, and products as life evolved before the Archean.

We highlight how special, if not exceptional, Earth is for its ability to remain inhabited over billions of years in the face of a brightening sun, cooling interior, emerging tectonics and continents, and broad shifts in ocean-atmosphere redox, among other first-order controls and consequences of life's beginnings. An implicit goal of our approach is that searches for life elsewhere within and outside our solar system should go beyond present and past habitability and consider whether a given planet or moon would have ever generated life. The Earth, with its tangible geological record, can serve as a model for studying exoplanets, their potential habitability, and biosignatures those worlds might produce.

In truth, we can only scratch this surface given how big the gaps are in our knowledge of life's beginnings on Earth, but it is a good time to start. We begin by considering the broad-scale context in which prebiotic chemistry might occur, including the stellar and planetary environment. These baseline factors may predetermine later challenges and opportunities as the planet and its system form and evolve.

**Planetary System Scale**

At the scale of the planetary system, there are several important factors contributing to habitability. Star type has been explored extensively as an independent variable when considering the best locations to look for habitable exoplanets. Since smaller stars have longer lives, and Earth's geobiologic record suggests that complex life took billions of years to arise, large stars with lifetimes of only 10s of millions of years are likely unsuitable hosts for highly evolved organisms, although possibilities of 'simple' microbial ecosystems cannot be ignored. Because planets around these larger, short-lived bright stars might only exist in a Hadean-like prebiotic state, smaller and longer-lived F, G, K, and M type host stars are often prioritized when looking for potentially habitable planets beyond our solar system (i.e., Kasting et al., 1993; Rugheimer et al., 2013; Kaltenegger and Lin, 2021). In the evolution of smaller G-type stars like our Sun, the pre-main sequence phase can be much more intense compared to the main sequence phase, which can result in an early stripping of a planet's surface volatiles (Luger and Barnes, 2015). Stellar age can also play a critical role, as the conditions of a star change during its lifetime. Perhaps most interesting and challenging in understanding Earth's early oceans is the so-called Faint Young Sun paradox, which attempts to reconcile the evidence for liquid water with the known faintness of the younger Sun in the early chapters of Earth history (Charnay et al., 2020), including the Archean. Analysis of this issue, and particularly relationships to $CO_2$ and methane as early modulators of global temperature, has furthered our understanding of early Earth generally—and more specifically by illuminating early atmospheric compositions and redox and ocean chemistry that favored or challenged potential prebiotic pathways. These relationships can be viewed in tandem with models and data for stellar evolution and the changing properties of our planet's interior.

To better understand the unique aspects of the Earth, as they may relate to our planet's ability to host life, it is key to also understand the broader planetary context in which similar planets may form and evolve. The activity of the star is critical because high-energy flares can destroy a planetary atmosphere and sterilize its surface. Recent studies suggest that M-dwarf flares, which are more frequent than those around other star types, can lead to ozone destruction and UV-C (ultraviolet-C) band penetration, particularly for magnetically unprotected planets (Tilley et al., 2019; France et al., 2020). Additionally, the extended pre-main-sequence phase of M dwarf stars may present issues for both habitability and remote characterization of an exoplanet through stripping surface volatiles and adding spectral uncertainty (Luger and Barnes, 2015; Godolt et al., 2019), although these concerns do not necessarily preclude habitability. A robust magnetic field could protect against flares, or life could originate in subaqueous or subterranean environments. Another possibility is that life might arise on an old M dwarf once it has entered a later stable phase (France et al., 2020). K dwarfs also exploit the observational advantage and may be more quiescent regarding stellar activity over long time scales, which could make them better places to host life (Arney, 2019). Given this range of outcomes for potentially habitable planets, the circumstances that conspired to make the Archean Earth habitable and capable of hosting biotic innovation are remarkable and perhaps rare (Lyons et al., 2024).

In addition to the radiative impact of the star, gravitational interactions between the host star and planet can also have significant influence on the planet. Beyond providing the right level of heating to support surface liquid water, the orbital location of the Earth provides an additional advantage in that the semi-major axis is sufficient to prevent tidal locking and orbital resonance. Close-in planets likely experience tidal locking, which can produce significantly uneven surface temperatures such as permanent magma oceans on the dayside of tidally locked planets orbiting M-type stars (Barnes, 2017; Pierrehumbert and Hammond, 2019). However, the resulting heat imbalances can be mitigated by oceanic and atmospheric heat transport (Hu and Yang, 2014; Yang et al., 2020). Further, the presence of companions, in the form of moons or neighboring planets, can induce orbital resonances which oppose tidal locking (Barnes, 2017). While Earth's orbit is not close-in enough to induce tidal locking, close-in planets present the best targets for remote observation given modern technology, both highlighting the necessity to understand Earth's history but also explore a broad range of potentially habitable exoplanets.

The overall makeup of the planetary system is also important. The presence of companions, such as the giant planets in our own solar system, can affect the planetary migration history. While the role of Jupiter and other gas giants is not fully understood, the influence of giant planets on system architecture and gravitational disturbances leading to impact events is thought to be significant (Horner and Chambers, 2010; Batygin and Laughlin, 2015; Grazier, 2016). Furthermore, the presence of so-called 'hot Jupiters' in exosolar systems fundamentally altered our understanding of the migratory history of the inner solar system and suggests that past histories of other planets may be more dynamic than can be inferred from current conditions. Gravitational perturbations during planetary migration can lead to changes in the impactor rate and exogenous delivery of volatile compounds from comets and asteroids (Strom et al., 2005; Fassett and Minton 2013). These impacts, as we discuss below, may have played an essential role in the initiation of life on Earth.

**Planet formation, composition, and moons**

In addition to the first-order influence of stellar size and evolution, planetary companions and impacts can shape the orbital properties of a planet, which in turn, affect the likelihood that life could evolve. The best explored orbital parameter relevant to habitability is the semi-major axis. Kopparapu et al. (2013) and other researchers studied this parameter to quantify the location of the liquid water 'habitable zone', which is often used as a first-order pass on constraining potential habitability (Figure 1). The idea of a "continuously habitable zone" or "habitable zone for complex life" builds on these initial ideas and begins to capture some of the complexity in long term habitability necessary to support the long, and ongoing, biological evolution observed on the Earth (Schwieterman et al., 2019; Tuchow and Wright, 2023). Other orbital parameters, such as the orbital obliquity and tilt, may have implications either for the ability of life to emerge or our ability to detect it because life can have disparate impacts over time on a tilted planet that may be detectable by remote observation (i.e., the Keeling curve) (Spiegel et al., 2009; Armstrong et al., 2014). The orbital properties can develop through gravitational interactions with other planets or through major impact events. The relatively temperate orbit of the Earth with low obliquity and eccentricity may also play a role in the ability of the planet to originate and sustain life.

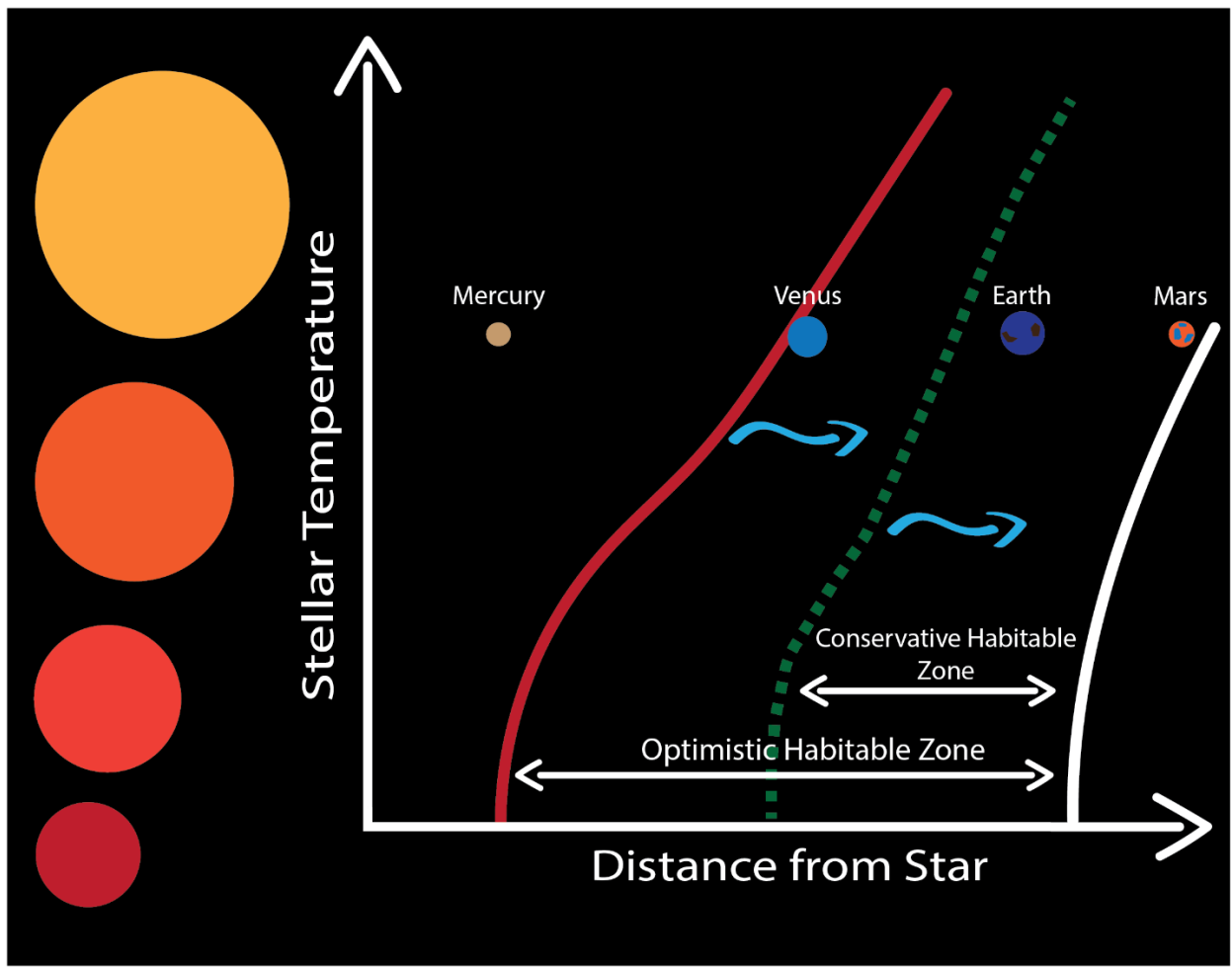

Figure 1: Schematic illustration, not to scale, of the distance of the habitable zone around a star compared to the relative temperature of the star. The habitable zone is defined as the region around a star where liquid surface water could be sustained, given a sufficient atmosphere. The optimistic habitable zone, between the red (leftmost) and white (rightmost) curved lines, is defined as between early Venus and early Mars when they had liquid water, although that possibility is decidedly less known for Venus. The conservative habitable zone, between the dashed green line and white curved line, does not include early Venus. The relative positions of Mercury, Venus, Earth, and Mars are plotted relative to the solar system's habitable zone. Blue curved arrows indicate the movement of the habitable zone away from the star as the star ages and increases in temperature.

Additional key factors in planetary conditions are the bulk composition and size, both of which are connected to the formation pathways and protoplanetary disc properties. Many of the topics covered later in this chapter, such as the geologic activity of a planet, presence of a carbon-silicate cycle, and outgassing of a secondary atmosphere, are connected to the bulk planetary composition, which is ultimately driven by the makeup of protoplanetary disc and evolutionary pathways therein. A critical compositional parameter is the initial water budget during planet formation, which is a first-order control on whether there could be oceans on the earliest Earth as evidenced in Hadean zircon oxygen isotope studies (Wilde et al., 2001). Some debate remains whether water was incorporated directly during accretion and or was delivered by impactors from the wetter outer solar system (King et al., 2010; O'Brien et al., 2018).

While the size constraints on habitability are not fully known, size is directly associated with cooling time, which in turn can have implications for both the presence of a magnetic field and the possibility of plate tectonics. In our own solar system, the warm, wet past of Mars offers tantalizing clues to the potential role of planetary size in long term habitability (see Kane et al., this volume). For our own planet, its core formation may have also paved the way for the appearance of the geodynamo, giving rise to the magnetic field. The magnetic field is believed to have played a key role in stabilizing the long-term prebiotic conditions by deflecting solar wind and retaining the atmosphere. Geochemical proxies suggest that the dynamo must have appeared by the early Archean (~3.5 billion years ago) (Tarduno et al., 2010), but it remains unclear if it was present in the Hadean.

Moon formation also affected the first stages of Earth's formation and initiation of a magnetic fields. During the first 10s of millions of years, Earth collided with the Mars-sized proto-planet, Theia resulting in the formation of the Moon. This collision potentially altered the chemical composition of the Earth, caused and stabilized the Earth's axial tilt, and, over the last 4.5 billion years, has slowed down the planet's rotation into the current 24-hour day. While the Earth's moon was formed via a large impact, moons can also result via capture such as Phobos and Deimos, the moons of Mars, but this process does not affect the composition and properties of the planet in the same way. Planets without moons may not experience tides (important for wet-dry cycles) or may be more likely to become tidally locked to their star, which has implications for heat transfer and long-term habitability (Barnes, 2017; Checlair et al., 2017; Pierrehumbert and Hammond, 2019)

**Origins of life**

There are wide-ranging views on how life began on Earth, such as the popular "genes first" (RNA World) perspective. It is not our intent to unpack the range of conceptual models that dominate conversations about the origins of life. Instead, we focus on environmental requirements and possibilities that often cross the boundaries of some or even all these models (such as the need for liquid water and sources of prebiotic precursor molecules). Our approach is thus best described as a feasibility study for given model requirements that illuminate the likelihood of a particular step or necessity based on what we know about early Earth—in ways that can also inform choices made in the designs of experimental simulations. In other words, we can now do a better job of putting boundary conditions on the possibilities for earliest Earth, including life-yielding pathways. We focus here on several of the most essential considerations.

*Atmospheric chemistry*

In general, the atmosphere would have played two main roles in the emergence of Earth's earliest life. First, the atmosphere can provide the precursor molecules or the "building blocks" for the formation of life through different prebiotic chemical pathways (e.g., Miller and Urey, 1959; Kobayashi et al., 2023; Wogan et al., 2023). Next, the atmosphere holds a critical role in providing appropriate surface condition in terms of temperature, radiation, total pressure, and more for the emergence and sustainability of life. For example, the greenhouse gas content of the atmosphere would have been key in maintaining clement surface conditions under the faint young sun (FYS) (Charnay et al., 2020). Our history reveals the importance of degassing and impact events on the formation and evolution of an early atmosphere.

A first-order control on atmospheric composition is the differentiation of the Earth's interior. Earth's core formation, based on the constraints of hafnium-tungsten isotope systematics, is believed to have occurred very early, within the first few 10s of Myr of planetary formation (reviewed in Carlson et al., 2014). This step profoundly influenced the oxidation state of the mantle and ultimately the atmosphere. However, no consensus has been reached on the oxidation state of the early mantle. Some papers argued for early magma and crust with reduced oxidation states (Yang et al., 2014), and thus the possibility of a reducing early atmosphere cannot be entirely ruled out, while others suggested an $fO_2$ that would have favored an oxidizing early atmosphere composed of $CO_2$-$N_2$ , though some species such as CO could be present in trace amounts and catalyze reactions (Hirschmann, 2012; Kasting, 2014; Sossi et al. 2020, Gaillard et al., 2022). This distinction has important implications for the atmospheric/subaerial generation and accumulation of building blocks for life, which may require a reducing environment (Benner et al., 2020; Wogan et al., 2023). If the prebiotic atmosphere were indeed oxidizing, we must consider the potential mechanisms of producing at least transient reduced conditions, such as $H_2$ production following impact events discussed below (Zahnle et al., 2020).

Much focus has been placed on the greenhouse gas composition of the early atmosphere, and many studies have attempted to reconcile the apparent necessity for greenhouse gases, such as $CH_4$, to offset the FYS and our understanding of the oxidation state of the atmosphere (Charnay et al., 2020). Prior to the origin of methanogenic organisms, abiotic systems would have been the primary source for atmospheric $CH_4$. Abiotic $CH_4$ can be produced from several geologic mechanisms: mainly via high-temperature magmatic processes occurring in volcanic and geothermal systems or by low-temperature gas-water-rock interactions (or post-magmatic processes) occurring in crustal serpentinization sites (Etiope and Sherwood Lollar, 2013). Recent work has shown the production of prebiotic molecules in volcanic plumes and the role of volcanic lightning in prebiotic chemistry (Hess et al., 2021; Bada, 2023). However, the composition of these volcanic gasses is largely determined by the quartz-fayalite-magnetite mineral buffer in the mantle, which would likely have led to a weakly oxidizing atmosphere composed of $CO_2$ and $H_2O$ (Trail et al., 2011; Zahnle et al., 2020). These weakly oxidizing atmospheres would present a challenge for many origin of life scenarios, although some studies suggest that in neutral conditions, inefficient production of prebiotic molecules remains possible, albeit slow to accumulate (Cleaves et al., 2008; Trainer, 2013). This possibility expands the changes needed in the overall atmosphere to plausibly produce prebiotic molecules such as amino acids.

It is hypothesized that a reducing atmosphere containing $CH_4$ and other hydrogen-containing compounds could provide a suitable environment for the synthesis of nucleic acid precursors from atmospheric gases and shield the photochemical products from UV radiation. In a reducing $CH_4$-rich early atmosphere, photolysis products of $CH_4$, methyl radical ($CH_3$) and methylene ($CH_2$), could react with atomic nitrogen to produce the nitrile hydrogen cyanide (HCN) (Tian et al., 2011).

HCN is suggested to be a key ingredient in prebiotic chemistry and may have acted as the source of carbon and nitrogen, which was required for synthesizing main building blocks of life such as amino acids, nucleotides, and lipids (Ferus et al., 2017; Kobayashi et al., 2023). While $CH_4$ could be critical in atmospheric prebiotic chemical scenarios, it may cause problems as well. Buildup of $CH_4$ and other hydrocarbons could form a haze, which, when dense enough, shields the planetary surface from UV radiation, causing cooling, although also protecting organic molecules (Arney et al., 2016). Hazes could also directly source life's building blocks, such as nucleotides, through photochemical reactions, but the preservation of these molecules in "warm little ponds" is likely dependent upon stable and ideal methane concentrations (Pearce et al., 2024). Importantly, the timing and pathway of $CH_4$ production is critical and likely changed from primarily potential inorganic reactions (serpentinization, metallic Fe reduction) to primarily biological activity with the evolution of methanogens from the Hadean to the Archean. The Archean advent of microbial methane production was a milestone, with profound implications for early climate (Lyons et al., 2024), which will be discussed in far greater detail elsewhere in this volume (e.g., Goldblatt et al. and Ebadirad et al., this volume).

Atmosphere loss is another key determinant for evolution over geologic time as atmospheres can retain heat and maintain surface pressure allowing liquid water (Hecht, 2002; Dong et al., 2017; Jakosky, 2021). Atmospheric retention depends on several factors including the presence or absence of a magnetic field, stellar type and activity, outgassing rates, and photochemical reactions. In our own solar system, we can see three examples of potentially habitable early planets—Venus, Earth, and Mars— though only Earth has maintained long-term habitable surface conditions (Kane et al., this volume). Venus has instead entered a runaway greenhouse effect (Westall et al., 2023), and Mars has lost an appreciable atmosphere (Lammer et al., 2013).

*Plate tectonics*

One significant milestone on a young terrestrial planet is the generation of crust, which evolves from mantle differentiation. There are several pathways to produce felsic rocks. First, giant impacts can result in large-scale partial melting of mafic rocks, yielding large-volume igneous differentiation (Johnson et al., 2018, 2022; Latypov et al., 2019). Felsic materials can also derive as a product of progressive fractional crystallization of a melt pool. Given the predicted high flux of meteorites in the Hadean, Johnson et al. (2018) argued that impact melting may have been the predominant mechanism that gave birth to early felsic crusts. Second, in a world without modern-style plate tectonics, multiple processes might have contributed to the mantle-crust differentiation, such as heat pipes, drips, and plumes, along with delamination and upwellings (Stern, 2018). Finally, modern-style plate tectonics—characterized by subduction-driven horizontal movement of blocks—stands out as the major pathway for crustal production. Nevertheless, the timing of the initiation of modern-style plate tectonics remains highly controversial, ranging from the Hadean (Korenaga, 2018, 2021a), the early Archean (e.g., ~3.8 Ga, Drabon et al., 2022; ~3.5 Ga, Greber et al., 2017), the Archean-Proterozoic transition (e.g., ~2.5 Ga, Lee et al., 2016), to the late Proterozoic (Stern, 2018, 2020).

Following the Moon-forming impact and resultant magma oceans, early water oceans likely existed perhaps dating back to 4.3 Ga (Mojzsis et al., 2001). However, they were likely relatively shallow (<1 km) due to water's greater solubility in magma (reviewed in Korenaga, 2021a, b). Importantly, however, the mantle's storage capacity for water is still debated, with estimates ranging from high to low levels, leaving open the possibility of relatively deep early oceans as well (Dong et al., 2021). Additionally, the evolution of surface water is tied to the development of plate tectonics as hydrated slabs are one of the main mechanisms that returns water to the mantle, affecting the potential depth and extent of oceans over early continents and topography. Scenarios with "fast" plate tectonics could lead to shallower oceans (Korenaga, 2021b); however, throughout the Hadean, plate tectonics likely would have slowed down as the mantle cooled from a mid-Hadean thermal peak (Kirkland et al., 2021) and differentiated, allowing the formation of deeper oceans. Simply, as the crust developed, the extent and depth of the oceans likely evolved as well. This possibility has implications for wet-dry and freeze-thaw cycles, as the depth and extent of the oceans would directly correlate with the amount of exposed landmass. Subaerial land may have been essential, because the products of atmospheric prebiotic chemical reactions could accumulate and undergo further polymerization. However, very few restraints exist on the volume of land mass that would be exposed, particularly due to uncertainties regarding the modes of crustal growth and evolution discussed below, as well as potential changes in the depth of the oceans through time (Korenaga, 2013).

The composition of the mantle likely changed through the Hadean as well. In general, the mantle transitioned from being homogenous before planetary differentiation to a chemically heterogenous, Mg-rich mantle characterized by the presence of Fe-rich blobs during the Hadean (Johnson et al., 2022). This process would have given rise to the generation of olivine-rich crust with high density. Moreover, higher temperatures and higher water contents would have contributed to the low viscosity of the mantle. Combined with the dense crust mentioned above, and the lubrication function of surface waters, the mantle's low viscosity could have allowed fast moving plate tectonics, seafloor resurfacing, and weathering processes over this period (Miyazaki and Korenaga, 2022). The unique olivine-rich composition of crust should facilitate hydrothermal serpentinization between the crust and ocean water, which can produce $H_2$ and, through reduction of $CO_2$, $CH_4$ that compensated for the oxidizing conditions stemming from early core formation (Ueda et al., 2021; Miyazaki and Korenaga, 2022).

Moreover, because the hot early core and mantle could have resulted in a higher core heat flux and thus a greater flux, mantle plumes, hotspot islands, and oceanic plateaus are expected to have been more abundant than present (reviewed in Korenaga, 2021a, b). Impact-generated topography, with an amplitude of a few kilometers (Marchi et al., 2014), could also have created stable patches of dryland essential for various prebiotically chemical reactions. As the mantle cooled and differentiated further, the composition and mechanics could have changed as well, potentially leading to (proto)subduction and partial melting (O'Neill and Debaille, 2014; O'Neill et al., 2017). However, komatiitic (Mg-rich) rocks are well recognized in the Archean, implying the continued presence of ultramafic and mafic rocks throughout the Hadean (Mole et al., 2014).

The evolution of continental crust, and more specifically the time-varying change in the extent of exposed land, is of importance for models of life's origin invoking wet-dry cycles (e.g., Lahav et al., 1978; Campbell et al., 2019). A conceptual model of these different possibilities is presented in Figure 2. These controls on the synthesis of proto-biopolymers, driven by the oscillating activity of water, could not have occurred if all the early continents were subaqueous. Additionally, the early mafic to ultramafic crustal composition could have facilitated serpentinization (Ueda et al., 2021), which is relevant to hydrothermal origin of life scenarios discussed below. Several modes of crustal development may have existed, or co-existed, during the Hadean, each with their own implications for prebiotic chemistry and the origin and evolution of life (Figure 2).

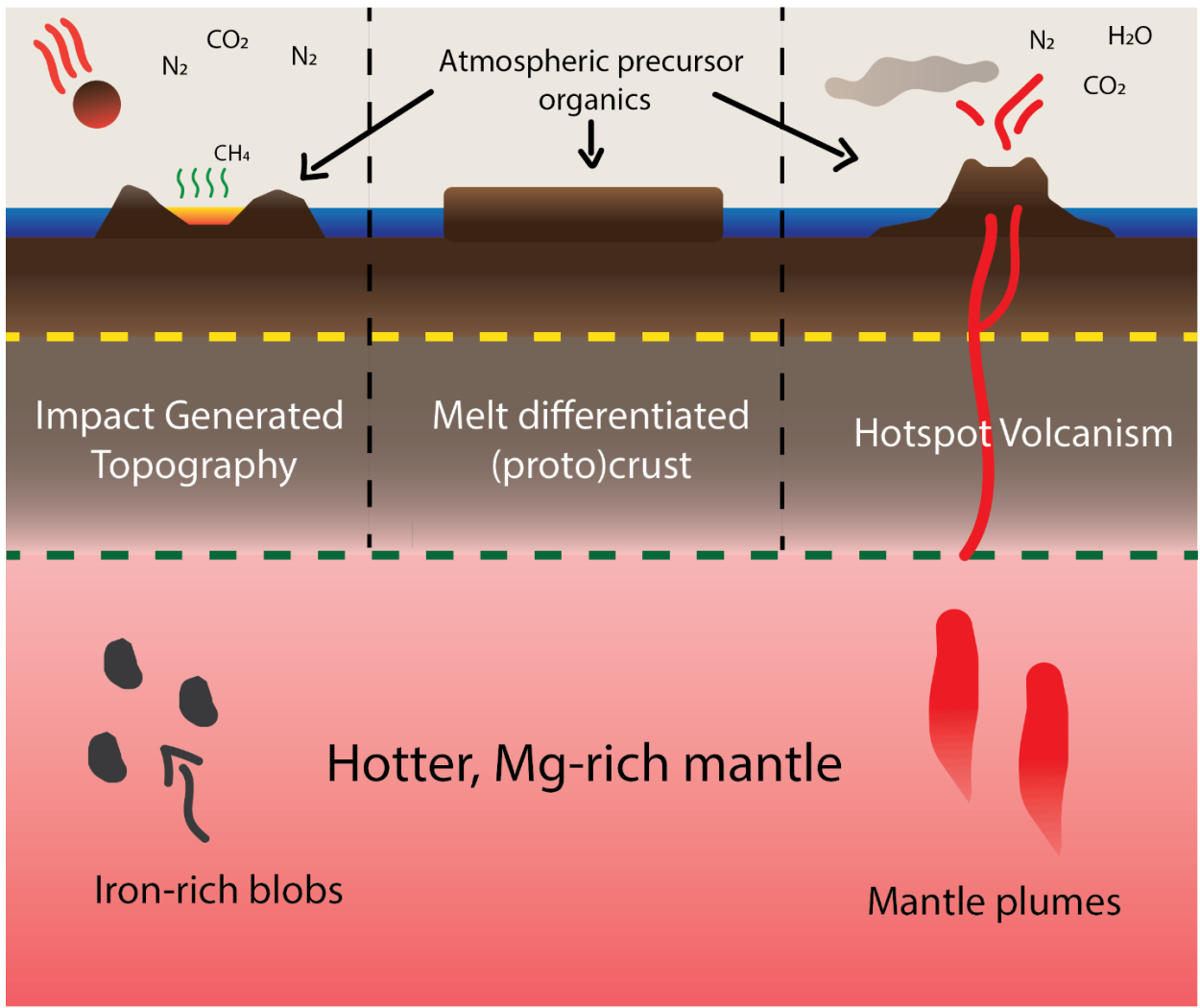

Figure 2: Illustration showing three different possibilities for mechanisms to generate exposed landmasses during the Hadean, all overlying a hotter, more Mg-rich mantle compared to today's. From left to right: impact generated topography (reviewed in Osinski et al., 2020), (proto)crust generated from differentiation during magma cooling (reviewed in Korenaga, 2021a), and hotspot volcanism related to mantle plumes (reviewed in Wright et al., 2024). Impact generated topography may also lead to transient reducing conditions due to metallic iron content in impactors, leading to degassing of methane ($CH_4$) in contrast to a relatively oxidized atmosphere of carbon dioxide ($CO_2$) and nitrogen ($N_2$) and water vapor ($H_2O$) that would be volcanically outgassed from an oxidized early mantle. Exposed land surfaces were potentially necessary to accumulate prebiotic chemical products of atmospheric chemistry. The mantle was likely hotter and more Mg-rich than today, also including iron-rich blobs as the Earth continued to differentiate. The blue dashed line represents the crust-mantle boundary, while the green dashed line represents the lithosphere-asthenosphere boundary.

*Wet-dry and freeze-thaw cycles*

Wet-dry cycles are non-enzymatic (abiological) processes by which catalytic and replicating polymers are synthesized. Models invoking wet-dry cycles require a host of prerequisite chemical and physical conditions on a prebiotic Earth, as outlined in Damer and Deamer (2020) and Deamer et al. (2022), including the requirement for exposed land surfaces, as discussed above. The dry phase of the wet-dry cycle is diffusion limited, which prompts the formation of new bonds between monomers (Higgs, 2016). During the wet phase, the monomers and polymers are redistributed, which increase the chances of collisions with other molecules during the next drying phase to form new bonds and thus oligomer formation (Higgs, 2016). Repeated wet-dry cycles (Figure 3) have been shown - in both laboratory experiments and theoretical models - to result in the formation of significantly longer polymers (e. g., oligoesters, oligopeptides, nucleotides, and oligonucleotides) when compared to prolonged dry heating or, on the other extreme, molecules residing persistently in solution without dramatic swings in water volume (Damer and Deamer, 2020; Forsythe et al., 2015; Pearce et al., 2017).

Wet-dry cycles include but are not limited to several potential prebiotic Earth environments where thermal evaporation, geyser activity, rainfall and seasonal flooding, tides, and/or deliquescence could occur (Damer and Deamer, 2020; Deamer et al., 2022; Frinkel-Pinter et al., 2022). Hotspot islands and seamounts, or mid-Hadean continents as suggested above, on an early Earth would create the opportunity for wet-dry cycles, although the existence of subaerial land, and the longevity of these wet-dry systems, is unclear (Becker et al., 2018; Campbell et al., 2019; Damer and Deamer, 2020; Deamer et al., 2022; Forsythe et al., 2015; Lahav et al., 1978). Additionally, the role of UV radiation, discussed in the atmospheric chemistry section above, is pertinent to subaerial wet-dry cycles as photons, both in the visible and UV range, have been shown to provide the free energy required for self-organization (Pascal et al., 2013). Likely there was a balance between the positive and negative effects of UV radiation, also related to atmospheric conditions and shielding of organics from UV photolysis (Arney et al., 2016).

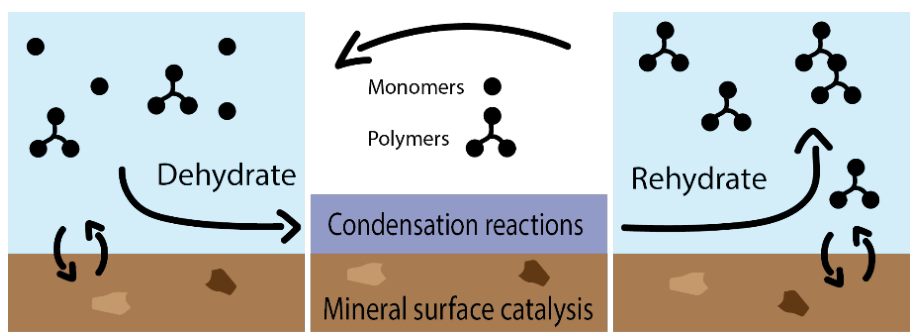

Figure 3: Cycle of dehydration and rehydration, which can occur via wet-dry, such as hot springs or shallow marine hydrothermal systems, or freeze-thaw cycles, causing condensation and polymerization reactions of prebiotic organics into more complex organic molecules. Monomers are indicated by individual black circles, while polymers are indicated by groups of black circles. Exchange between rock surfaces, where minerals can catalyze reactions, is indicated by circular arrows at the fluid-rock interface.

The essential organic building blocks used to construct polymers and oligomers via wet-dry cycles could have been amino acids (Damer and Deamer, 2020). Indeed, amino acids were present and were likely even abundant on the prebiotic Earth through either endogenous formation; delivery in meteorites (Miller and Urey, 1959; Kvenvolden et al., 1970; Glavin et al., 2020); or via reactions between primary precursors, including hydrogen cyanide (HCN), cyanamide ($H_2NCN$), cyanoacetylene (HCCCN), cyanogen (NCCN), ammonia ($NH_3$), and cyanic acid (HCNO) (Benner et al., 2020). As discussed in the atmospheric chemistry section above, the redox state of the Hadean atmosphere is controversial, but it would have controlled whether reduced molecules could have formed. Several pathways to reducing atmospheres are possible, including $CH_4$ degassing from melt pools and transient reducing atmospheres following bolide impacts, but their longevity and ability to create prebiotic precursor molecules is not well constrained (Wogan et al., 2023).

Similar to wet-dry cycles, freeze-thaw cycles can cause polymerization reactions (Marín-Yaseli et al., 2020; Menor-Salvan and Marín-Yaseli, 2013) and could have relevance to life's origins on moons in the solar system with thick ice layers over vast liquid oceans (e.g., Enceladus and Europa; Menor-Salvan and Marín-Yaseli, 2012; Menor-Salvan and Marín-Yaseli, 2013), perhaps in combination with photochemistry by UV radiation that played a role in building larger, more complex organics.

The importance of sea ice on early Earth, beyond possible freeze-thaw pathways that could stimulate polymerization, would include consequences for surface albedo and thus warming scenarios that could have favored liquid oceans. For example, a higher early salinity due to the absence of continental cratons to sequester halite deposits (Knauth, 2005; Ueda et al., 2016) would reduce the production of sea ice at high latitudes and thus lower the net surface albedo and decrease the $pCO_2$ level necessary to prevent a snowball-Earth scenario (Olson et al., 2022). This relationship could have extended into the early Archean as continents developed or continued to evolve and emerge above sea level.

*Hydrothermal systems*

The composition of the Hadean crust was significantly more mafic compared to the modern and may have been limited to basaltic to komatiitic seafloor (O'Neil et al., 2011; Nebel et al., 2014; Herzberg, 2016), especially if large-scale subaerial continents had not yet formed from crustal differentiation related to plate-tectonic processes. This situation would have given rise to the generation of olivine-rich crust with high density. The unique olivine-rich composition of crust should facilitate serpentinization reactions between the oceanic crust and ocean water (Miyazaki and Korenaga, 2022).

The early ocean may have been characterized by circumneutral to basic pH with carbonate alkalinity generated by weathering of impact ejecta or volcanic material (Kadoya et al., 2020). Alternatively, the high $pCO_2$ levels predicted, particularly in the absence of appreciable persistent warming by methane (Zahnle et al., 2020), might have driven the oceans toward pH values closer to 6 (Halevy and Bachan, 2017). Importantly, many of the seafloor weathering and hydrothermal reactions are pH- and temperature-dependent, thus changes in these parameters affect potential prebiotic chemical pathways (Kempe and Degens, 1985; Sleep and Zahnle, 2001; Zahnle and Sleep, 2002; Kempe and Kazmierczak, 2002). Moreover, the contrast between pH values of fluids released from alkaline seafloor vents and the ambient seawater (estimated as acidic, neutral, or basic) would have controlled the proton motive force that could have been essential to initiating and sustaining life under such conditions (Martin et al., 2008; Cockell et al., 2016). In contrast, acidic hydrothermal vents with low pH (2-3) may still provide a proton gradient, but the higher temperatures are generally less conducive to continued abiotic organic synthesis (Martin et al., 2008). Additionally, the speciation of carbon is also pH dependent and plays a role in reduction reactions in hydrothermal systems (Seewald et al., 2006). Given remaining uncertainties, the range of possible gradients between vents and seawater spans many pH units — a debate that carries implications for other chemical species that may have been important in prebiotic processes, such as borate, which can stabilize molecules such as ribose (Ricardo et al., 2004).

Serpentinization—the reaction of water with Fe- and Mg-rich minerals, such as olivine and pyroxene in ultramafic rocks (in this case oceanic crust)—produces $H_2$ in an exothermic reaction (Proskurowski et al., 2008) (Figure 4). These warm, reducing fluids can then migrate upwards through the crust and vent at the seafloor. The dissolved $H_2$ can reduce dissolved $CO_2$ via reactions on metal-rich minerals to reduced carbon species such as $CH_4$, thus providing a substrate for prebiotic chemical reactions (reviewed in Schrenk et al., 2013). The leap from prebiotic chemistry to biochemistry begins with the synthesis of organic molecules. Afterwards, these molecules must be polymerized and encapsulated within semipermeable membranes. Monnard and Deamer (2002) describe how polymerization and self-assembly of organic molecules is inhibited in solutions of high ionic strength, such as the modern ocean. On the other hand, Jordan et al. (2019) investigate a broader range of organic molecules produced by Fischer-Tropsch reactions (abiotic assembly of hydrocarbons using metal catalysts) and concluded that the presence of longer chain fatty acids, and the inclusion of 1-alkanols, widen the range of pH, temperature, salinity, and ionic strength conditions that would allow spontaneous vesicle formation to encompass hydrothermal conditions expected at alkaline hydrothermal vents. Furthermore, ammonia can also be generated during hydrothermal alteration of komatiitic rocks by iron-mediated thermochemical reduction of nitrogen ($N_2$), nitrite, and nitrate (Brandes et al., 1998; Nishizawa et al., 2021).

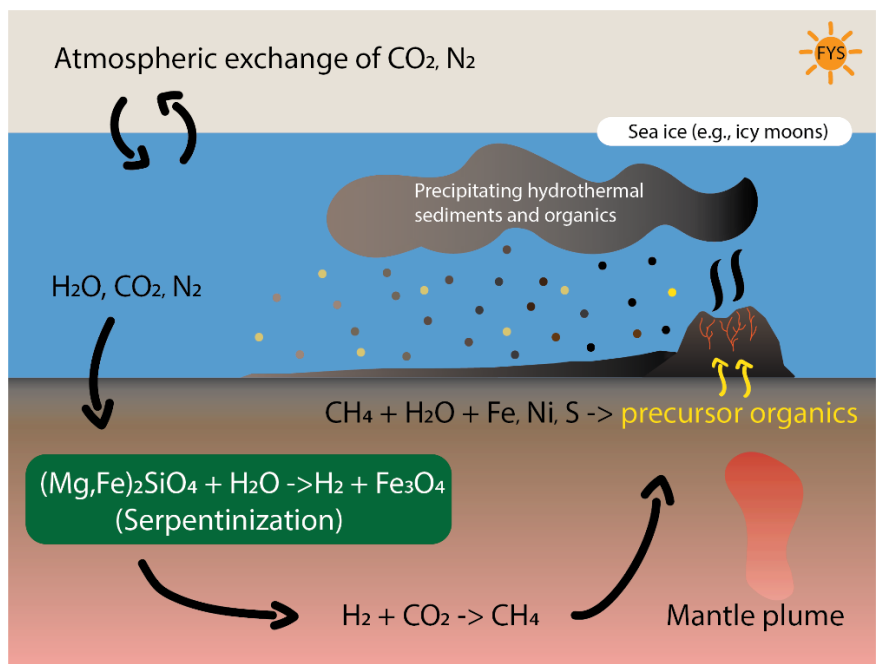

Figure 4: Illustration of potential Hadean hydrothermal vent systems and prebiotic chemical pathways. As seawater, containing carbon dioxide ($CO_2$) and nitrogen ($N_2$), flows through mafic crust, serpentinization can occur (green box). Magnesium (Mg)- and iron (Fe)-rich minerals react with seawater, generating hydrogen ($H_2$), which can reduce $CO_2$ to methane ($CH_4$). Additional reactions with Fe, nickel (Ni), and sulfur (S) can result in more complex organics that are vented back into seawater through hydrothermal vents. Hydrothermal vents may be essential on worlds with ice shells, affecting atmospheric-ocean exchange.

Based on clues provided from DNA, the last universal common ancestor may have been living in a geochemically active site filled with $H_2$, $CO_2$, and Fe, possibly in a hydrothermal setting (Weiss et al., 2016; Wimmer et al., 2021). Because of this, some workers have placed emphasis on the role of Fe, nickel (Ni), and sulfur (S) minerals/alloys that could catalyze anabolic reactions of reduced carbon species along temperature and pH gradients that would have favored the proton motive force along electrochemical gradients at the boundaries between the venting fluids and seawater (Amend and Shock 1998; Seyfried and Foustoukos, 2007; Sojo et al., 2014; Sojo et al., 2016; Lane, 2017; Li et al., 2018; Omran et al., 2020). These boundaries may have been analogs for cell membranes, with the Fe, Ni, and S alloys resembling enzymes such as ferredoxin used in cellular processes such as the acetyl-coA-pathway (Varma et al., 2018; Jordan et al., 2019). Furthermore, the substrates available for abiotic reactions via alteration of ultramafic rocks are comparable to those used by life today: $CO_2$, $CH_4$, and water ($H_2O$), in contrast to the molecules (e.g., HCN) invoked for other prebiotic chemistries such as the Miller-Urey experiment (Sojo et al., 2016; Harrison and Lane, 2018).

Polymerization of molecules such as RNA has been less intensely studied in hydrothermal settings because of the preservational challenges that come with hydrolysis, leading many to favor prebiotic chemical pathways that require dry conditions or wet-dry/freeze-thaw cycles with episodes of low water activity. We discussed above how these conditions can also favor condensation reactions that result in production of larger organic molecules from smaller precursors. An interesting combination of these processes is subaerial or shallow marine hydrothermal systems that could have wet-dry cycling induced by tides or other processes. In these environments, prebiotic compounds and organics, such as those discussed above from serpentinization reactions, could be sourced from hydrothermal systems and undergo polymerization reactions, such as in tidal pools refilled by hydrothermal brines (Barge and Price, 2022; Baumgartner et al., 2024).

However, Preiner et al. (2018) point out that low water activity environments can form in aqueous environments when water is consumed during serpentinization, dissolved gases react with the crust, and hydrophobic molecules undergo phase separation, all of which could create localized, low water activity environments. This possibility is supported by Baross and Martin (2015), who argued that accumulation of hydrophobic molecules, not dissimilar to the "tar" often produced in prebiotic chemistry experiments including Miller and Urey (1959), mimics the role that many enzymes play in cellular functions by excluding water from active sites.

Hydrothermal fields, though not individual vents, can remain active for 10s to 100s of thousands of years (Kelley et al., 2005), though they are variable in composition, redox, and temperature through both space and time (Haase et al., 2007; Lang et al., 2013). While metal availability in seawater and on the continents is typically controlled by atmospheric oxygenation and weathering patterns (i.e., molybdenum liberation during oxidative weathering of the continents), hydrothermal vents source their elements primarily from water-rock interactions in crustal rocks or overlying sedimentary cover (Fisher, 1998; McCollom and Shock, 1998). Consistent with these possibilities, there is evidence for hydrothermally altered and serpentinized sediments in the early Archean age Barberton Greenstone Belt of South Africa (Westall et al., 2018).

*The impact of impacts*

Our current understanding of the Hadean Earth is limited by the lack of preserved geological material. However, lunar crater histories and solar system models allow us to consider the role bolide impacts may have played in Earth's early planetary and biotic evolutionary patterns and parameters. Impacts could have affected Earth through delivery of volatiles such as water, light elements, reactive phosphorus, and organic compounds in addition to influencing atmospheric composition and generating topography, among many other possible factors.

Many studies have investigated the role of impactors in Earth's volatile budget, particularly the delivery of water to the early Earth. Impactors may have additionally delivered other compounds (i.e., organics) and elements (evidenced by the higher than expected abundance of highly siderophile elements in the crust and mantle today). Calculations show that up to 0.5% ((0.7–3.0) x $10^{22}$ kg) of Earth's mass may have been delivered post-formation by continued impacts (Marchi et al., 2014). For example, schreibersite, a reactive P bearing mineral, is relatively common on meteorites but is exceedingly rare on Earth, where P is instead hosted in minerals such as apatite (Walton et al., 2021). Phosphorus is a common limiting nutrient in modern Earth environments and was likely even rarer on the early Earth (Walton et al., 2021), and the limited abiotic reactivity of phosphate has presented a challenge in models of prebiotic chemistry (Pasek et al., 2017).

Therefore, impact events could have supplied reactive P to the early Earth, although other pathways for making P available include photochemical reduction of P in the presence of sulfide (Ritson et al., 2020), metamorphic reactions between phosphate and $H_2$ or ferrous iron (Pasek et al., 2022; Baidya et al., 2024), or the generation of phosphide minerals (such as schreibersite) by lightning strikes or association with gypsum (Benner et al., 2020; Hess et al., 2021).

The possible contributions by impacts to the probiotic potential of an early Earth include (1) direct delivery of organic molecules, (2) changing the redox state of the pre-existing atmosphere from oxidized to reduced conditions, and (3) providing shock energy for prebiotic synthesis (Fegley et al., 1986; Chyba and Sagan 1992; Goldman and Tamblyn, 2013; Lee et al., 2022). As a function of size, composition, and velocity of the impactor, post-impact atmospheres can range from fully reduced with a completely vaporized ocean to localized effects that lead to transient reduced conditions and spatially limited impact topography (e.g., Zahnle et al., 2020). Briefly, impactors can deliver reduced compounds or elements (in particular metallic Fe that leads to $H_2$ production) that can lead ultimately to reduced conditions in the atmosphere and/or enhance release of reduced compounds within the target from impact-generated melt pools (Marchi et al., 2014; Zahnle et al., 2020; Marchi et al., 2021; Wogan et al., 2023). A resulting $CH_4$-rich atmosphere might have helped offset the FYS (as discussed above), although high levels of $CH_4$ would lead to organic haze production and attendant climate cooling (Zahnle et al., 2020), and these transient effects would have occurred on time scales that varied with the nature of the impactor and the collision and the frequency of events.

A fundamental result of impacts is the generation of topography (Figure 5), including crater rims and central highlands. These topographic highs could provide dry land prior to the onset of formation and emergence of continents and could thus support wet-dry cycles. Further, Osinski et al. (2020) discuss how hydrothermal vents can form because of energy released during impact, also providing locations for potential prebiotic reactions and habitability. Impact events forming large craters can mix ocean and mantle material, leading to exchange between the surface environment and deeper solid Earth, including bioessential elements (Marchi et al., 2014). In some cases, impact events may even directly catalyze the formation of prebiotic molecules, such as amino acids, by coupling the reduction of $H_2O$, $CO_2$, and $N_2$ in the ocean and atmosphere to the oxidation of metallic Fe, Ni, and organics delivered by the impactor (Osinski et al., 2020; Takeuchi et al., 2020).

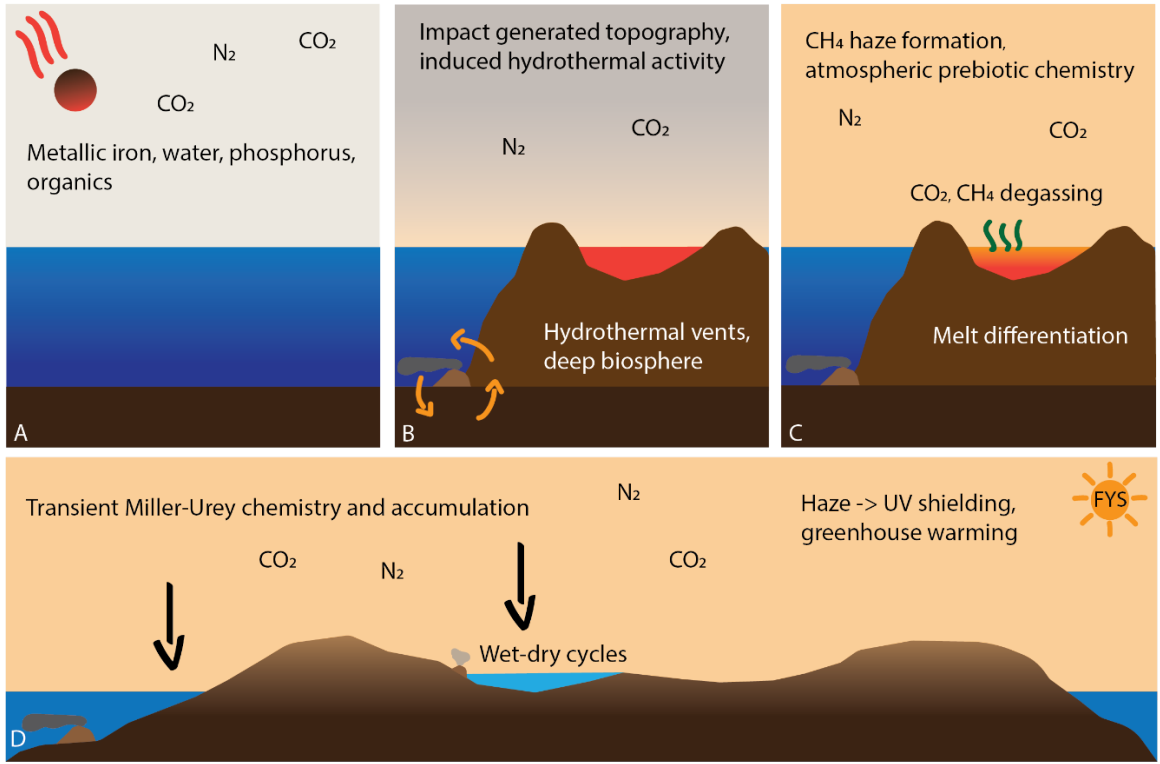

Figure 5: Schematic illustrating the role that impacts may have had on the early Earth in sequence from Panel A – Panel D. Panel A illustrates an incoming impactor. Panel B shows the resulting crater topography, induced hydrothermal activity, and dust in the atmosphere. Panel C indicates the potential for methane ($CH_4$) formation and melt differentiation. Panel D zooms in on exposed land and the potential for wet-dry cycles and accumulation of atmospheric precursor molecules. Impact events may have delivered reduced compounds such as organics or metallic iron, reactive phosphorus, and water to the Earth. Reducing power from impacts could lead to transient reducing atmospheres and organic haze formation, both allowing Miller-Urey style atmospheric chemistry and greenhouse warming that helped compensate for the faint young sun (FYS). Impacts could generate topography, stimulate hydrothermal activity, and create exposed land surfaces, which may be necessary for various prebiotic chemical pathways.

Impact events were traditionally viewed as a detriment to the origin and sustainment of life on the early Earth (Maher and Stevenson, 1988). Several models based on cratering of the Moon and Mercury (e.g., Marchi et al., 2009; Morbidelli et al., 2012; Marchi et al., 2013) indicate an uptick in impact events around 4.1 Gya related to a dynamical instability in the giant planets (Bottke et al., 2012). However, recent studies call into question the intensity and sterilizing effects of large impacts during the Hadean (e.g., Grimm and Marchi, 2018), and the severity and even existence of the late heavy bombardment is debated (Bottke and Norman, 2017). These estimates raise questions about the likelihood of sterilization at the transition between the Hadean and the Archean, suggesting that life that originated during the Hadean could have continued without interruption into the Archean. Moreover, the zone of habitability on a planet extends into its subsurface, crust, and sedimentary cover (Abramov and Mojzsis, 2009), further supporting the possibility of early life surviving into the Archean. The persistence of life into the Archean, with or without the LHB or impacts in general, would set the stage for advanced metabolisms and proliferation, leading to truly global effects on the Earth system ranging from changing ocean chemistry to atmospheric oxygenation (Lyons et al., 2024).

**Conclusions**

The early Earth likely had liquid water, but the presence of continents, their composition, and extent is debated. This uncertainty raises important questions about the various pathways proposed for prebiotic organic synthesis and eventually the origin of life. For example, without continents, could wet-dry cycles have occurred? Conversely, on icy moons or exoplanets such as water worlds, are hydrothermal vents the only prebiotic pathway? While freeze-thaw cycles may span this gap, some of the proposed chemical reactions require photochemistry and/or photochemical products under methane rich atmospheres. Likely, many of the discussed processes and reactions were likely occurring simultaneously on the Hadean Earth, and the crossover zones between these separate environments were likely key in the origin and evolution of life. An exciting possibility for these complex, crossover environments could come from continued impacts through the Hadean. These may have delivered necessary volatiles or elements to the earth surface, stimulated hydrothermal activity, generated topography, or directly catalyzed prebiotic reactions. The impacts of impacts on atmospheric composition likely continued into the Archean (Marchi et al., 2021).

Regardless of the exact pathway, life had taken hold by the end of the Hadean and began to leave evidence in the geologic record of the Archean by at least 3.7 billion years ago, specifically suggestions of biological carbon fixation. It is our hope that the details provided here will clarify how early life and habitability might have initiated and persisted into the Archean—in other words, this is the preface to the environments and life that followed. In doing so, we also argue that lessons learned from early Earth should and do guide us in our efforts to find and explore other habitable planets and moons, within and beyond our solar system. Perhaps most remarkable, as we dive headfirst into the Archean, is that Earth maintained it habitability, including oceans over most of our history, and fostered the beginnings and proliferation of microbial life—in response to and as a driver of environmental change. And Earth did so despite changes as dramatic as a brightening sun, cooling interior, the first appearance of continents, and first-order changes in the composition of our atmosphere. The roadmap provided by the Hadean, despite the many still-fuzzy details, gives us a lot to build on as we think about what followed and why—on Earth and perhaps beyond.